\documentclass[reprint,aps,prl,superscriptaddress,noeprint]{revtex4-2}

\usepackage[normalem]{ulem}
\usepackage{siunitx}
\usepackage{amsmath}
\usepackage{amsbsy}
\usepackage{amssymb} 
\usepackage{graphicx}
\usepackage{float}
\usepackage{bm}
\usepackage[caption=false]{subfig}
\usepackage{hyperref}
\hypersetup{
	colorlinks=true,
	linkcolor=blue,
	citecolor=green,
	filecolor=magenta,
	urlcolor=blue
}
\usepackage{siunitx}
\usepackage[dvipsnames]{xcolor}

\begin{document}
\title{Enhancing spin squeezing using soft-core interactions}	

\author{Jeremy T. Young}
\email[Corresponding author: ]{jeremy.young@colorado.edu}
\affiliation{JILA, University of Colorado and National Institute of Standards and Technology, and Department of Physics, University of Colorado, Boulder, Colorado 80309, USA}
\affiliation{Center for Theory of Quantum Matter, University of Colorado, Boulder, Colorado 80309, USA}

\author{Sean R. Muleady}
\affiliation{JILA, University of Colorado and National Institute of Standards and Technology, and Department of Physics, University of Colorado, Boulder, Colorado 80309, USA}
\affiliation{Center for Theory of Quantum Matter, University of Colorado, Boulder, Colorado 80309, USA}

\author{Michael A. Perlin}
\affiliation{JILA, University of Colorado and National Institute of Standards and Technology, and Department of Physics, University of Colorado, Boulder, Colorado 80309, USA}
\affiliation{Center for Theory of Quantum Matter, University of Colorado, Boulder, Colorado 80309, USA}
\author{Adam M. Kaufman}	
\affiliation{JILA, University of Colorado and National Institute of Standards and Technology, and Department of Physics, University of Colorado, Boulder, Colorado 80309, USA}

\author{Ana Maria Rey}
\affiliation{JILA, University of Colorado and National Institute of Standards and Technology, and Department of Physics, University of Colorado, Boulder, Colorado 80309, USA}
\affiliation{Center for Theory of Quantum Matter, University of Colorado, Boulder, Colorado 80309, USA}

\date{\today}

\begin{abstract}
We propose a new protocol for preparing spin squeezed states in controllable atomic, molecular, and optical systems, with particular relevance to emerging optical clock platforms compatible with Rydberg interactions. By combining a short-ranged, soft-core potential with an external drive, we can transform naturally emerging Ising interactions into an XX spin model while  opening  a many-body gap. The gap helps maintain the system  within a  collective manifold of states where metrologically useful spin squeezing  can be generated at a level   comparable to the spin squeezing  generated in systems with genuine all-to-all interactions. We examine the robustness of our protocol to experimentally-relevant decoherence and show favorable performance over typical protocols lacking gap protection.

\end{abstract}


\maketitle

As the means to control quantum systems has progressed in recent decades, so too has the ability to create and harness quantum entanglement for improved quantum technology. 
In the context of quantum sensors, this entails applying entangled resources to increasingly push beyond the standard quantum limit (SQL) --- the fundamental noise floor for uncorrelated particles --- towards the fundamental limits imposed by quantum mechanics \cite{Giovannetti2006,Giovannetti2011,Toth2014, Szczykulska2016,Pezze2018}. 
Although current state-of-the-art optical clocks provide some of the most precise measurements in physics \cite{Bothwell2022,Zheng2022,Brewer2019},
they will eventually reach a point where improvements in sensing capabilities based on uncorrelated atoms have diminishing returns due to both fundamental physical and practical constraints. In light of this, the utilization of entanglement provides an additional axis for optimization, which will be crucial for the next generations of optical clocks once the limits of these constraints are reached. 

In recent years, tweezer arrays of neutral atoms have emerged as a promising new platform for optical clocks \cite{Madjarov2019, Norcia2019, AYoung2020}, driven by a number of recent advances, including the rapid preparation of tunable arrays with high filling fractions and single-atom control \cite{Lee2016,Barredo2016,Endres2016, Norcia2018,Browaeys2020,Bluvstein2021} and half-minute-scale coherence times on optical clock transitions \cite{Norcia2019, AYoung2020}. Such platforms combine the control and high-duty cycles of ion clocks \cite{Brewer2019,Chou2010,Schmidt2005} with the scalability of optical lattice clocks \cite{Bothwell2022,Zheng2022} while mitigating their respective drawbacks, such as interatomic collisions in lattice clocks or large shot noise in ion clocks.
Moreover, in these systems tunable Ising  interactions via Rydberg states \cite{Saffman2010, Saffman2016, Wu2021, Morgado2021}  that decay as $1/r^\alpha$ with interparticle distance $r$ offer a natural avenue for the generating metrologically useful entanglement  in the form of spin squeezing \cite{Kitagawa1993,Wineland1992,Wineland1994}.  However, as long as the dimension of the array $D \le \alpha$, such interactions yield spin squeezed states that provide only a small, constant noise reduction that is independent of particle number \cite{Foss-Feig2016}. 

\begin{figure}[t]
    \centering
    \includegraphics[scale=.56]{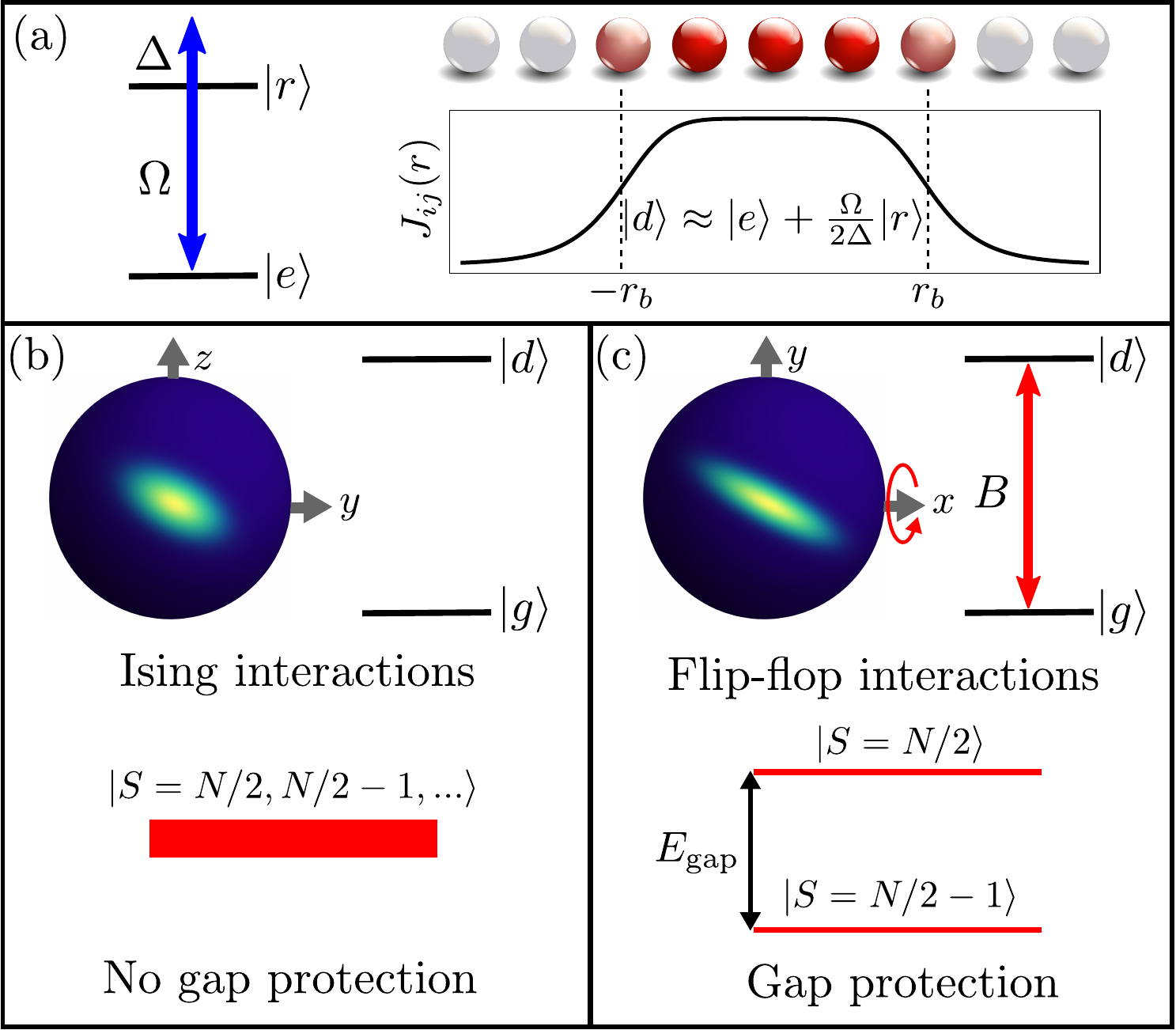}
    \caption{Comparison of two approaches to generating spin squeezing using dressed Rydberg interactions. (a) The state $|e\rangle$ is weakly dressed with a Rydberg state $|r\rangle$ with Rabi frequency $\Omega$ and detuning $\Delta$, resulting in a soft-core potential with blockade radius (i.e., range) $r_b$ for the dressed state $|d\rangle$. (b) Forming a two-level system with $|d\rangle$ and a ground state $|g\rangle$, realizes an approximate OAT Hamiltonian for $N \lesssim N_b$. However, for larger systems, the perturbation to OAT quickly limits any additional gains with increasing $N$ due to the absence of a gap. (c) By applying a strong transverse field $B$ via a drive between $|g\rangle$ and $|d\rangle$, the system realizes an approximate OAT Hamiltonian with an additional term which opens a gap $E_\text{gap}$ between different $S$ manifolds. Due to the resulting gap protection, OAT scaling is preserved well beyond $N \sim N_b$, leading to enhanced squeezing.}
    \label{fig:schematic}
\end{figure}

Rydberg dressing has been proposed  as a way  to modify the form of the interaction via a dipole blockade mechanism, resulting in a tunable soft-core potential, which can be turned off by extinguishing the dressing field \cite{Pupillo2010,Johnson2010, Henkel2010, Jau2016, Arias2019,Zeiher2016,Zeiher2017, Borish2020,Guardado-Sanchez2021, Schine2021}. This enables effective collective interactions  within the region set by the soft-core potential, which can improve squeezing but with minimal gain for systems larger than the potential range \cite{Gil2014}. This  limits the utility of the generated Ising interactions given that the soft core radius in 
current many-body experiments has been $\sim 1$-2 times the inter-atomic spacing \cite{Zeiher2016, Zeiher2017, Borish2020, Guardado-Sanchez2021}, partly to avoid   the onset of avalanche processes that are caused by blackbody radiation \cite{Goldschmidt2016,Aman2016,Boulier2017}.
Although Rydberg dressing can be implemented in a variety of neutral atom platforms, we shall focus on tweezer arrays, which provide means of controlling both the atomic density and number, both of which can be used to mitigate blackbody avalanche processes and, as we shall show, are beneficial for optimizing the performance of our protocol.

Here, we propose to combine the collective effects of a soft-core potential and
a strong transverse field via a coherent drive to convert the  Ising model to an XX model  \cite{Jurcevic2014,Richerme2014,Wall2017,Kiely2018,Friis2018,Monroe2021}. The XX model features a many-body  gap  that protects  the manifold of collective states and mitigates the effect of non-collective interactions \cite{Rey2014,Smale2019,He2019,Perlin2020,Perlin2022} (see Fig.~\ref{fig:schematic}). 
We show that the interplay between  soft core interactions plus gap  protection dramatically extends the system sizes for which the optimal spin squeezing mimics that of a fully collective Ising model, also known as the one axis twisting (OAT) model  \cite{Kitagawa1993,Ma2011}. For example, in a 2D system of $32 \times 32$ atoms, a soft-core potential range of only 3 times the lattice spacing is needed to realize near-OAT squeezing, even though the number of atoms that fall within the soft-core potential is about 36 times smaller than the system size. In addition, we show that such collective behavior remains robust to the presence of fundamental decoherence processes for realistic tweezer array experiments.

\textit{Model}.---We consider a scheme where an internal  state $|e\rangle$ is dressed with a Rydberg state $|r\rangle$ via a drive with Rabi frequency $\Omega$ and detuning $\Delta$ \cite{Pupillo2010,Johnson2010,Henkel2010, Jau2016,Zeiher2016,Zeiher2017, Arias2019, Borish2020,Guardado-Sanchez2021, Schine2021}. The resulting dressed state $|d\rangle \approx |e\rangle + \frac{\Omega}{2 \Delta} |r\rangle$ and a ground state $|g\rangle$ are used to form an effective spin-1/2 system governed by a  Hamiltonian of the form
\begin{subequations}
\begin{equation}
    H = \sum_{i < j } J_{ij} (1/2+s_i^z) (1/2+s_j^z),
    \label{eq:Ising}
\end{equation}
\begin{equation}
    J_{ij} = \frac{\Omega^4}{8 \Delta^3} \frac{1}{1+(r/r_b)^6}, \quad \frac{C_6}{r_b^6} = -2 \Delta,
\end{equation}
\end{subequations}
where $s_i^\mu \equiv \sigma_i^\mu/2$ denote the spin-1/2 operators at site $i$, $J_{ij}$ is a soft-core potential with a range of blockade radius $r_b$ and $1/r^6$ tail, and $C_6/r^6$ is the van der Waals (vdW) interaction.
Physically, we can understand the emergence of this Hamiltonian as follows: at large distances, the Rydberg states interact weakly, leading to a vdW tail with reduced strength $f^2 C_6$, where $f \equiv \Omega^2/4 \Delta^2$ is the Rydberg fraction. However, at short distances where $|C_6/r^6| \gg |2 \Delta|$ (i.e., $r \ll r_b$), the excitation of more than one Rydberg atom is strongly suppressed due to blockade. As a result, the corresponding contribution to the light shift is suppressed, leading to a plateau of strength $J_0 \equiv 2 \Delta f^2 = \Omega^4/8 \Delta^3$. 
Finally, we note that in addition to the Ising interactions, an inhomogeneous longitudinal field is also introduced. Unless otherwise noted, we shall assume that these terms can be neglected either via spin-echo or a rotating wave approximation (RWA) in the presence of a strong drive as discussed below \cite{suppcite}.

In the system under consideration, an effective transverse field along the  $x$-direction can be generated by applying a drive which couples $|g\rangle$ and $|d\rangle$ with Rabi frequency $B$. 
In the limit of $B \gg (N-1) \overline{J} \equiv \frac{1}{N} \sum_{i,j} J_{ij}$, 
where $(N-1) \overline{J}$ is the average interaction each atom feels, and in the frame of the applied transverse field,  the Ising interactions  take the form of flip-flop interactions since under the RWA, the fast oscillating terms can be dropped out. The 
 final Hamiltonian takes  the form of an XX model \cite{Jurcevic2014,Richerme2014,Wall2017,Kiely2018,Friis2018,Monroe2021}
\begin{equation}
    H_\text{RWA} = \frac{1}{2}\sum_{i < j } J_{ij} (s_i^y s_j^y + s_i^z s_j^z).\label{eq:H_RWA}
\end{equation}
Note that in the course of making the RWA, the overall strength of the interactions have been reduced by a factor of two, indicating that 
the dynamics
will occur at a slower rate.
In the Supplement, we discuss the effects of a finite transverse field $B$ \cite{suppcite}.

\textit{Enhanced squeezing}.---For a system of $N$ spin-1/2 particles,  the Wineland spin squeezing parameter, $\xi$,  defined as   \cite{Wineland1992,Wineland1994}
\begin{equation}
    \xi^2 \equiv \frac{N \min{ \langle \Delta S_{\perp}^2 \rangle}}{|\langle \mathbf{S} \rangle |^2},
\end{equation} quantifies the reduction in the phase uncertainty beyond the SQL of $1/\sqrt{N}$.  Here $\mathbf{S} \equiv \sum_i \mathbf{s}_i$, and 
$\min{ \langle \Delta S_{\perp}^2 \rangle}$ denotes the minimum variance in directions perpendicular to the Bloch vector.
Note also that in contrast to OAT where the initial state  is typically oriented along $x$, for the above XX model squeezing dynamics happen when one starts   along $z$, and in this case the  spin squeezing will be  on  the $xy$-plane. 

For Ising interactions, the soft-core potential from weak Rydberg dressing allows for an improvement in squeezing over pure power-law interactions ~\cite{Gil2014}. This is because  within a blockade radius, the interactions are all-to-all, and thus the model realizes an effective OAT Hamiltonian $H_\text{OAT} \equiv \frac{\overline{J}}{2} S_z^2$ when the length of the system $N \lesssim N_b$. The optimal spin squeezing accessible via  OAT dynamics scales as,  $\xi^2 \sim N^{-2/3}$, in a time $\overline{J} t_{\text{opt}} \sim N^{-2/3}$ \cite{Kitagawa1993,Ma2011}.
However, as we increase the system size $N \gtrsim N_b$, the deviations from OAT can quickly become significant by coupling $|S,m_z\rangle, |S',m_z\rangle$ states ($S$ denotes the total spin and $m_\alpha$ the projection onto $S_\alpha$), which are degenerate in $H_\text{OAT}$. Accordingly, increasing the system size leads to limited squeezing improvement. For vdW interactions the  $1/r^6$ tail does allow for a moderate  enhancement  over the naive estimate of $N_b^{-2/3}$ based on OAT scaling, and in the thermodynamic limit, $\xi^2_\infty \propto r_b^{-.76 D} \propto N_b^{-.76}$ for $D \leq 3$ \cite{Gil2014}, where $N_b$ is the number of atoms within a blockade radius and the $\infty$ subscript denotes the thermodynamic limit.

To understand how the squeezing behavior changes for the XX model, it is convenient to re-express the Hamiltonian as
\begin{subequations}
\begin{equation}
    H_\text{RWA} = \frac{1}{2} H_\text{gOAT} + \frac{1}{2} \sum_{i < j} (\overline{J} - J_{ij}) s_i^x s_j^x,
\end{equation}
\begin{equation}
    H_\text{gOAT} = \sum_{i<j} J_{ij} \mathbf{s}_i \cdot \mathbf{s}_j - \frac{\overline{J}}{2} S_x^2.
\end{equation}
\end{subequations}
Here, we see that the effective OAT Hamiltonian has an additional SU(2) symmetric term. Although this is not a collective term, it nevertheless commutes with $\mathbf{S}^2$. As a result, this term will not couple different $S$ manifolds, but it will break their degeneracy in the OAT model, leading to a gapped OAT Hamiltonian $H_\text{gOAT}$. Like with the Ising model, the XX model will similarly lead to OAT for $N \lesssim N_b$. However,  the presence of a gap between different $S$ manifolds permits that as $N$ is increased beyond $N_b$, the deviations from $H_\text{gOAT}$ can be initially treated as a perturbation, extending the effective OAT behavior to larger $N$ compared to Ising interactions and providing enhanced squeezing. While a similar argument can apply for power-law interactions \cite{Perlin2020}, the soft-core potential here ensures that this perturbation grows much more slowly initially as the system size is increased.

\begin{figure}[t]
    \centering
    \includegraphics[scale=.56]{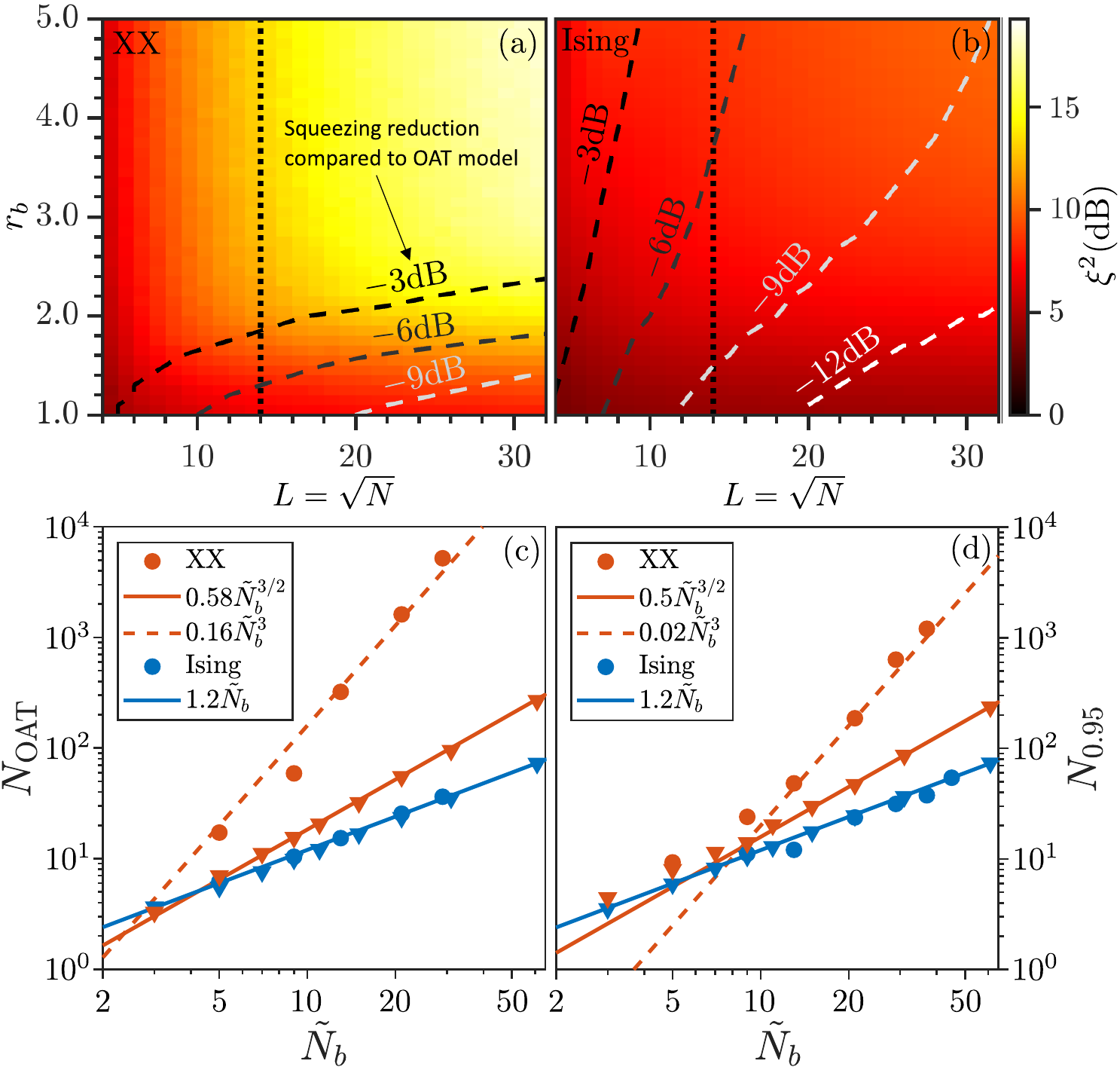}
    \caption{(Top) Comparison of the spin squeezing generated by (a) the XX model and (b) the Ising model. We show the optimal squeezing generated as a function of the blockade radius $r_b$ and the system side length $L$ for various two-dimensional systems of size $N = L\times L$. The black dotted line correspond to systems with $L=14$, for which we later consider the effects of decoherence. The dashed contours denote the reduction in the optimal squeezing for each model compared to that of an OAT model for the corresponding particle number (color variation selected for visibility).
    (c) Scaling of effective OAT atom number $N_{\text{OAT}}$ associated with $\xi^2_\infty$ for 1D (triangles) and 2D (circles) for a potential with a sharp cutoff as a function of $\tilde{N}_b \equiv N_b + 1$. 
    (d) Scaling of $N_{0.95}$ ($N$ at which $\langle S^2 \rangle/(N/2(N/2+1)) = 0.95$ at $t_\text{opt}$) as a function of $\tilde{N}_b$. Lines are meant to illustrate the scaling and are not fits.}
    \label{fig:squeezing}
\end{figure}

While the gap protection indicates that the XX model will realize improved squeezing compared to the Ising model, it need not necessarily be significant. To determine the degree of enhancement, we study both models numerically. For the Ising model, this can be done exactly. However, for the XX model, this is no longer possible and we must rely on numerical approximations. Here, we will consider using the discrete truncated Wigner approximation (DTWA) \cite{Schachenmayer2015a,Schachenmayer2015b,Zhu2019}, which shows good agreement with results using the time-dependent variational principle for matrix product states in 1D \cite{suppcite}; analogous benchmarks in 2D for spin systems with power law interactions exhibit similar agreement \cite{Muleady2022}. 

In Fig.~\ref{fig:squeezing}(a,b), we compare the squeezing performance of the resulting XX model vs.~the original Ising interaction using DTWA for vdW interactions. The squeezing for the XX model retains OAT scaling well beyond the naive expectation of $N \sim N_b \approx \pi r_b^2$, while the squeezing for the Ising interactions saturate at much smaller system sizes in comparison. For example, for $r_b=3$, corresponding to $N_b \approx 28$, the squeezing from Ising interactions begins to diverge from OAT around $N=9$. In contrast, for the XX interactions, the squeezing is only slightly reduced from OAT at $N=1024 \approx 36 N_b$. 

To understand the scaling of the squeezing with $N_b$, we define $N_{\text{OAT}}$ as the number of atoms necessary for OAT to realize $\xi^2_\infty$, thus determining the system sizes for which OAT scaling persists. We also investigate the gap protection by investigating the behavior of $\langle S^2\rangle /(N/2(N/2+1))$, which provides a measure of how collective the system is. In particular, we identify $N_{0.95}$, the number of atoms at which $\langle S^2\rangle /(N/2(N/2+1)) = 0.95$ at $t_{\text{opt}}$. This scaling is presented in Fig.~\ref{fig:squeezing}(c,d) for 1D and 2D with a sharp cutoff in the soft-core potential and periodic boundary conditions.
We see that both indicate that OAT scaling for the Ising model persists to $N \propto N_b$ and is independent of the dimension, as expected. In contrast, OAT scaling for the XX model persists to $N \propto N_b^{3 D/2}$, corresponding to $\xi_\infty^2 \propto N_b^{-D}$. Aside from the enhancement over Ising interactions, we see that the gap protection appears to be stronger at higher dimensions, leading to a further enhancement in the OAT scaling. 

Finally, let us discuss the behavior of the squeezing time. When $N \lesssim N_b$, $\overline{J}$ is approximately equal to the nearest-neighbor interaction strength $J_0$, so the squeezing time scales like $J_0 t_\text{opt} \approx \overline{J} t_\text{opt} \sim N^{-2/3}$. However, for $N>N_b$, we have $\overline{J} \approx J_0 N_b/N$, and the squeezing time scales like $J_0 t_\text{opt} \sim N^{1/3}/N_b$, leading to a tradeoff between enhanced squeezing and squeezing times, which can become particularly important in the presence of decoherence. 

\textit{Decoherence}.---While under ideal conditions we have shown the XX model outperforms  the Ising model, it remains to be seen whether this advantage can be realized in the presence of  relevant  decoherence processes found in experiments. There are two key distinctions regarding the effects of decoherence in the XX model as compared to the Ising model. First, the XX model is realized in a rotating frame, in which the dissipation takes on a different form. This can be understood by noting that due to the strong transverse field, each spin oscillates between being in the weakly-dressed Rydberg state $|d\rangle$, which decays, and being in $| g\rangle$, which does  not.  Second, the time scale necessary to realize the optimal squeezing is much longer for the XX model, owing to both the factor of two reduction in the interaction strength in the XX model relative to the Ising model and the comparatively longer time (scaled by the interaction strength) it takes to realize collective squeezed states in the XX model. As such, the XX model will generically be more affected by decoherence.

In the context of Rydberg dressed  atoms, the dominant form of dissipation will come from the Rydberg state $|r\rangle$ or from  $|e\rangle$. For the Rydberg decay, there are two scenarios we consider: decay to $|g\rangle$ ($\gamma_{rg}$) and decay to $|e\rangle$ ($\gamma_{re}$). In the case of the former, this will correspond to dissipation from the weakly dressed Rydberg state to $|g\rangle$ at rate $f \gamma_{rg}$;
for the latter, this will correspond to an effective dephasing of rate $f \gamma_{re}$. For dissipation from $|e\rangle$ at rate $\gamma_{eg}$, this will correspond to decay from the weakly dressed Rydberg state to $|g\rangle$ at rate $(1-f) \gamma_{eg}$. 

In the resulting effective spin-1/2 system, we include the effects of all three forms of decoherence via the Lindblad master equation
\begin{subequations}
\label{eq:master}
\begin{equation}
    \dot{\rho} = -i [H, \rho] + \sum_\mu\gamma_\mu \mathcal{D}_\mu[\rho],
\end{equation}
\begin{equation}
    \mathcal{D}_\mu[\rho] \equiv \sum_i \left[l_{\mu,i} \rho l_{\mu,i}^\dagger - \frac{1}{2} \{\rho,l_{\mu,i}^\dagger l_{\mu,i} \} \right],
\end{equation}
\end{subequations}
where $\mathcal{D}_\mu[\rho]$ describes a Lindbladian evolution term with rate $\gamma_\mu$ and Lindblad jump operator $l_\mu$. In the effective two-level system, there is decay at rate $\gamma_- = f \gamma_{rg} + (1-f) \gamma_{eg}$ with  $l_- = s^-$ and dephasing at rate $\gamma_d = f \gamma_{re}$ with  $l_d \equiv n_i = 1/2+s_i^z$. In the rotating frame, the system dephases in the transverse field direction at  a rate $\gamma_-$ and in the two orthogonal directions at a rate $(\gamma_- + \gamma_d)/2$ with  Lindblad jump  operators $s_x$ and $s_{y,z}$, respectively \cite{suppcite}.

For Ising interactions, it is  possible to solve Eq.~(\ref{eq:master}) exactly  \cite{Foss-Feig2013}. For the XX model, we use a dissipative generalization of 
DTWA \cite{Huber2022,Singh2022}. Briefly, this amounts to including fluctuations due to dissipation approximately via stochastic noise terms \cite{suppcite}. 

\begin{figure}[t]
    \centering
    \includegraphics[scale=.4]{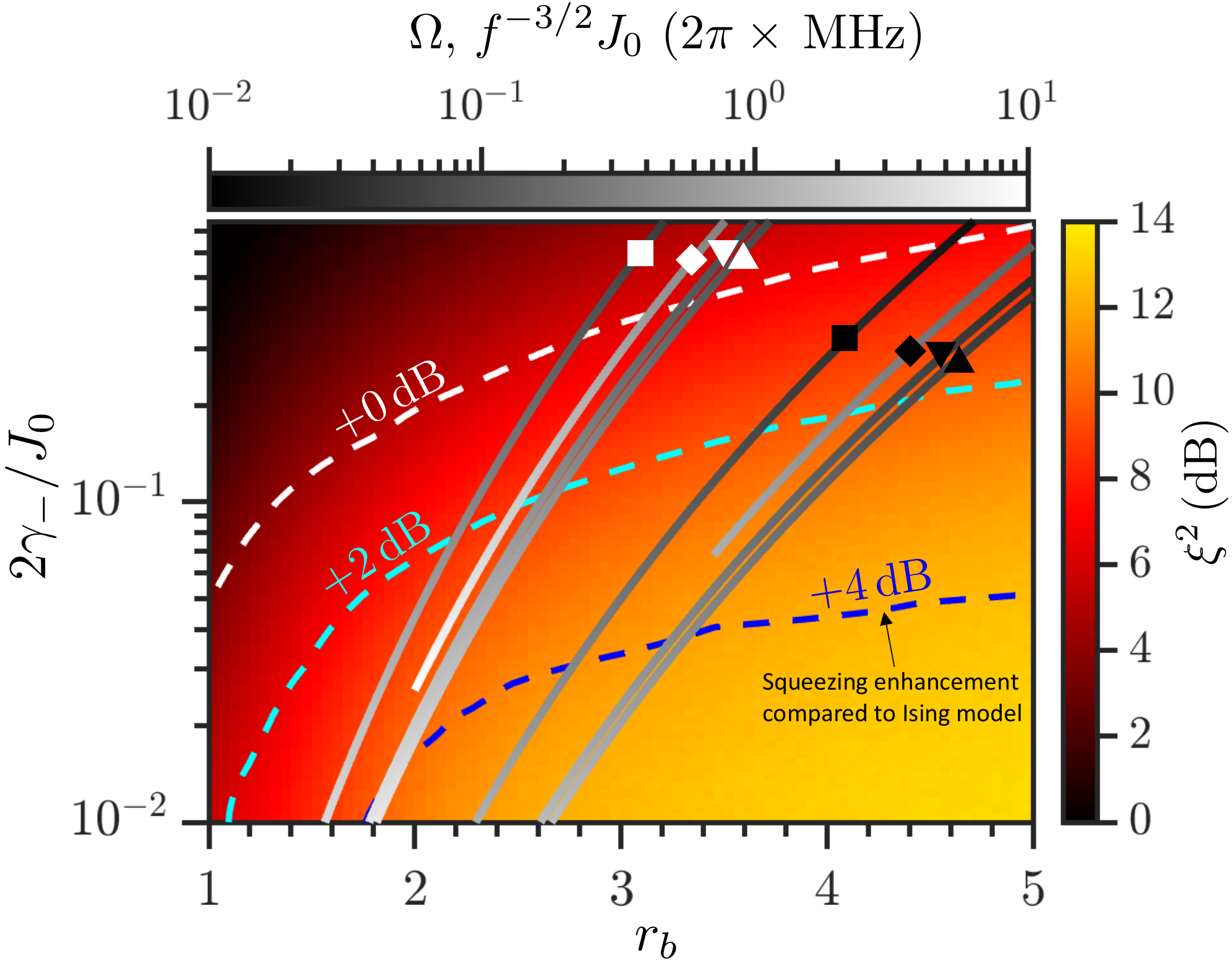}
    \caption{Spin squeezing generated by the XX model when incoherent effects are taken into account with $\gamma_- = \gamma_d = f\gamma_r/2$ in a $14 \times 14$ lattice. The dashed contours indicate the squeezing enhancement of the XX model over the Ising model, with decoherence taken into account in both models (color variation selected for visibility). We further illustrate lines of expected parameters for several possible systems \cite{suppcite}: $^{88}$Sr with $n=80$ (square), $^{88}$Sr with $n=41$ (diamond), $^{87}$Rb with $n=60$ (upside-down triangle), and $^{133}$Cs with $n=60$ (triangle). The lines labeled with black (white) symbols correspond to a fixed Rydberg fraction $f= 0.01$ ($f=0.001$), while the color scale associated with each line indicates the corresponding Rabi frequency $\Omega$, which can be related to the nearest-neighbor coupling via $J_0 = f^{3/2}\Omega$. We terminate each line when the required transverse field $B$ or dressing field $\Omega$ exceed typical thresholds for each atom \cite{suppcite}. (Squeezing color scale is the same as in Fig.~\ref{fig:squeezing})}
    \label{fig:exptsqueezing}
\end{figure}

In Fig.~\ref{fig:exptsqueezing}, we compare the performance of the Ising and XX models in the presence of dissipation in a $14 \times 14$ lattice. The relative values of $\gamma_-$ and $\gamma_d$ will depend on the choice of Rydberg state, branching ratios, and the temperature of the system. For simplicity, we take $\gamma_- = \gamma_d = f \gamma_r/2$, where $\gamma_r$ is the total decay rate of $|r\rangle$ at $T = \SI{300}{K}$ to all states; a more complete treatment would take into account branching ratios and losses to states outside the manifold we consider but this likely affects both Ising and XX implementations in a similar way. In the figure, we show the achievable spin squeezing in  the XX model as a function of   $\gamma_-/J_0$ and $r_b$ for several different choices of Rydberg atoms, excitation fraction, and Rydberg states (for further details, see \cite{suppcite}).  

We observe that increasing the blockade radius improves the overall squeezing generated by the XX model for any fixed $\gamma_-/J_0$ and also extends the relative advantage in performance of the XX model over the Ising model to stronger decoherence rates. This is largely a consequence of the fact that larger $r_b$ shortens the squeezing time, mitigating the effect of decoherence on the performance advantage of XX model over the comparatively faster Ising protocol.
We also note that increasing $\Omega$ for fixed $f$ (tracing the solid lines towards the bottom left of the plot) increases the interaction strength relative to the decoherence, but at the cost of requiring a comparatively larger transverse field for the RWA to remain valid.
Furthermore, we observe that by working at larger values of $f$, it is generally possible to attain smaller values of $\gamma_-/J_0$ for fixed $r_b$, thus improving the amount of generated squeezing. However, the associated faster timescales again raise the value of the requisite transverse field \cite{suppcite}, while sufficiently large $f$ may also lead to a breakdown of the weak dressing model.
Lastly, as we can observe from the $n=41$ and $n=80$ results for $^{88}$Sr and $f=0.001$ in Fig.~\ref{fig:exptsqueezing} (lines denoted by the white diamond and square, respectively), increasing $n$ for fixed $r_b$ and $f$ yields a slight degradation in the amount of squeezing generated, owing to a larger value of $\gamma_-/J_0$. Nonetheless, doing so also significantly reduces the overall value of $J_0$, again enabling one to operate with a much smaller transverse field.
Overall, we find that the XX model outperforms the Ising model for a wide range of experimental parameters and atoms, paving the way for generically realizing gap-protected enhanced squeezing in Rydberg platforms. 

\emph{Outlook}.---Although we have focused on spin squeezing with  Rydberg atoms, the driving idea discussed here   can potentially be used in other systems with finite-range interactions. For example, since even an interaction range of two sites is sufficient to realize significant enhancements in the squeezing, circuit-QED systems with interactions beyond nearest-neighbor may benefit from this approach \cite{Song2019,Groszkowski2022}. 
From a theoretical point of view, a comprehensive examination of the  various scaling behaviors with $r_b$ and how they depend  on the dimension, power-law tail, and system size,  as well as any potential connection between the scaling  with the presence Anderson's tower of states \cite{Comparin2022}, would be very illuminating. Additionally, the work here provides a foundation for developing more sophisticated protocols including  Floquet engineering  \cite{Borish2020, Vandersypen2005, Choi2020, Zhou2020, Geier2021, Scholl2022} or variational algorithms \cite{Kaubruegger2019,Kaubruegger2021, Marciniak2022} which might take further advantage of the combination of a soft-core potential with gap protection and  generate even better  and more robust  spin squeezing.

\begin{acknowledgments}
We thank P.~Rabl, J.~Huber, and T.~Roscilde for helpful discussions as well as W.~F.~McGrew and D.~Wellnitz for a careful reading and comments on the manuscript. This work is supported by the AFOSR grant FA9550-19-1-0275, by the NSF JILA-PFC PHY-1734006, QLCI-OMA-2016244, by the U.S. Department of Energy, Office of Science, National Quantum Information Science Research Centers Quantum Systems Accelerator, and by NIST. J.~T.~Y.~was supported in part by the NIST NRC Postdoctoral Research Associateship Award.
\end{acknowledgments}


\end{document}


\title{Supplemental Material: Enhancing spin squeezing using soft-core interactions}	

\author{Jeremy T. Young}
\affiliation{JILA, University of Colorado and National Institute of Standards and Technology, and Department of Physics, University of Colorado, Boulder, Colorado 80309, USA}
\affiliation{Center for Theory of Quantum Matter, University of Colorado, Boulder, Colorado 80309, USA}

\author{Sean R. Muleady}
\affiliation{JILA, University of Colorado and National Institute of Standards and Technology, and Department of Physics, University of Colorado, Boulder, Colorado 80309, USA}
\affiliation{Center for Theory of Quantum Matter, University of Colorado, Boulder, Colorado 80309, USA}

\author{Michael A. Perlin}
\affiliation{JILA, University of Colorado and National Institute of Standards and Technology, and Department of Physics, University of Colorado, Boulder, Colorado 80309, USA}
\affiliation{Center for Theory of Quantum Matter, University of Colorado, Boulder, Colorado 80309, USA}

\author{Adam M. Kaufman}	
\affiliation{JILA, University of Colorado and National Institute of Standards and Technology, and Department of Physics, University of Colorado, Boulder, Colorado 80309, USA}

\author{Ana Maria Rey}
\affiliation{JILA, University of Colorado and National Institute of Standards and Technology, and Department of Physics, University of Colorado, Boulder, Colorado 80309, USA}
\affiliation{Center for Theory of Quantum Matter, University of Colorado, Boulder, Colorado 80309, USA}

\date{\today}

\maketitle
\onecolumngrid

\renewcommand{\theequation}{S\arabic{equation}}
\renewcommand{\thesubsection}{S\arabic{subsection}}
\renewcommand{\thesubsubsection}{\Alph{subsubsection}}

\renewcommand{\bibnumfmt}[1]{[S#1]}
\renewcommand{\citenumfont}[1]{S#1} 

\pagenumbering{arabic}

\makeatletter
\renewcommand{\thefigure}{S\@arabic\c@figure}
\renewcommand \thetable{S\@arabic\c@table}

This supplemental material is organized as follows: in Sec.~\ref{transverse}, we investigate the effect that a finite transverse field has on the squeezing dynamics. In Sec.~\ref{bench}, we benchmark DTWA in 1D by comparing it to exact numerics using matrix product states (MPS) via the time-dependent variational principle (TDVP). In Sec.~\ref{exp}, we discuss the experimental parameters used in Fig.~3 of the main text and how they scale with the principal quantum number. Finally, in Sec.~\ref{DissDTWA}, we discuss the implementation of dissipative DTWA.

\section{Finite transverse field}
\label{transverse}
In this section, we investigate the effects that a finite transverse field has on the validity of the effective XX model
\begin{equation}
H = \sum_{i < j } J_{ij} (1/2+s_i^z) (1/2+s_j^z) + B \sum_i s_i^x.
\end{equation}
Note that aside from the desired interactions, there is also a longitudinal field $B^\parallel_i \equiv \sum_j J_{ij}/2$ introduced by the Rydberg dressing. Although in the bulk of the system this is homogeneous, near the boundaries, this is no longer the case. In the limit of $B \gg N \overline{J}$, we also have $B \gg B^\parallel$, so the longitudinal field is dropped via the RWA. However, for finite transverse field, it is important to take into account its presence.

There are two possible approaches to reducing the effects of the longitudinal field. In the first, a $\pi$ pulse is applied halfway through the evolution, effectively flipping the sign of the longitudinal field while leaving the interactions and transverse field unchanged. As a result, the evolution from the longitudinal field is removed in a spin-echo fashion. In the second, we can detune the drive used to generate the transverse field by the average $\frac{1}{N} \sum_i B_i^\parallel$. Since the longitudinal field is not homogeneous, this will not fully remove it, but it will drastically reduce its effect. Here, we will focus on the second approach.

\begin{figure}[h!]
    \centering
    \includegraphics[scale=.4]{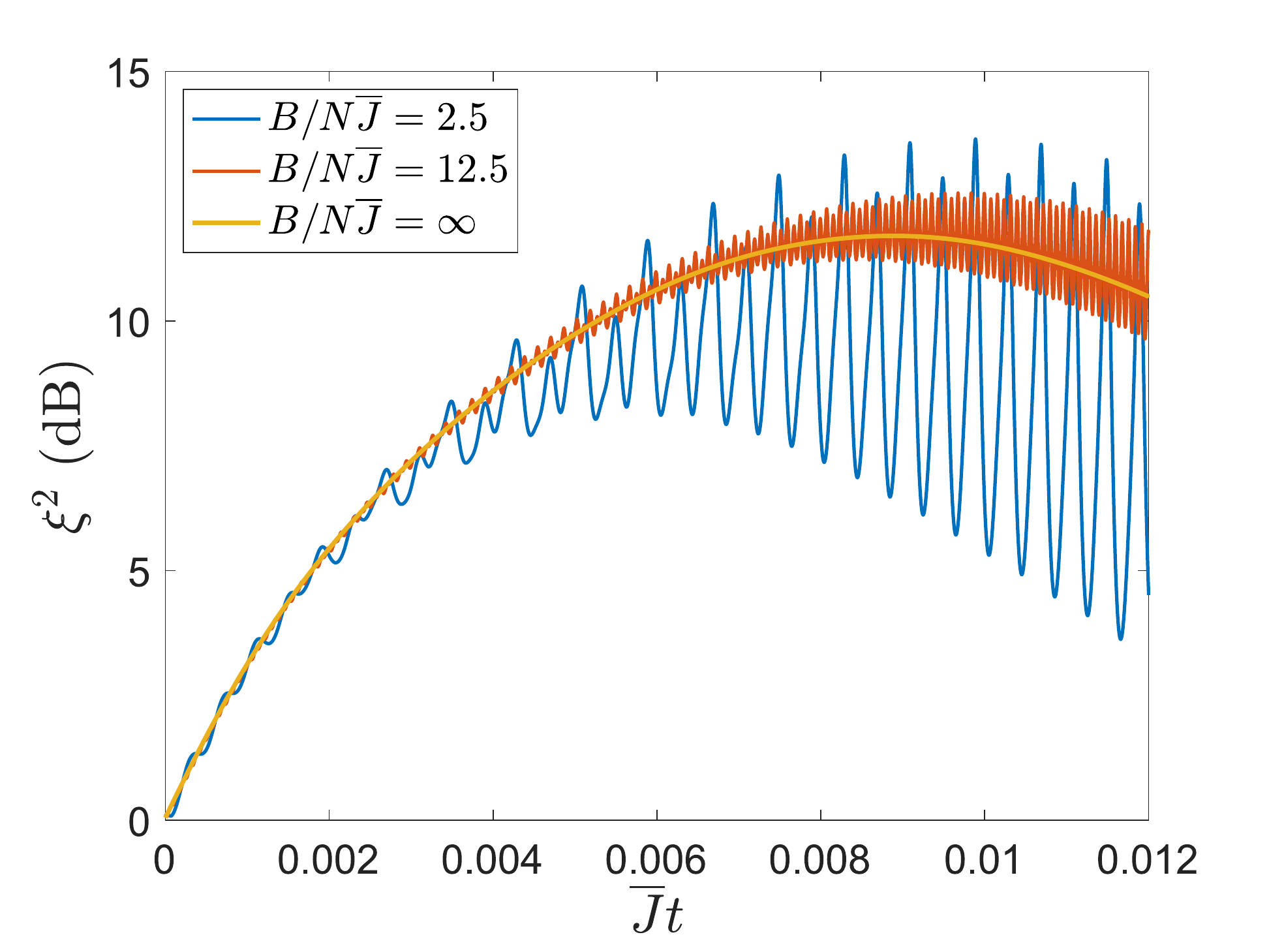} \qquad \qquad \includegraphics[scale=.4]{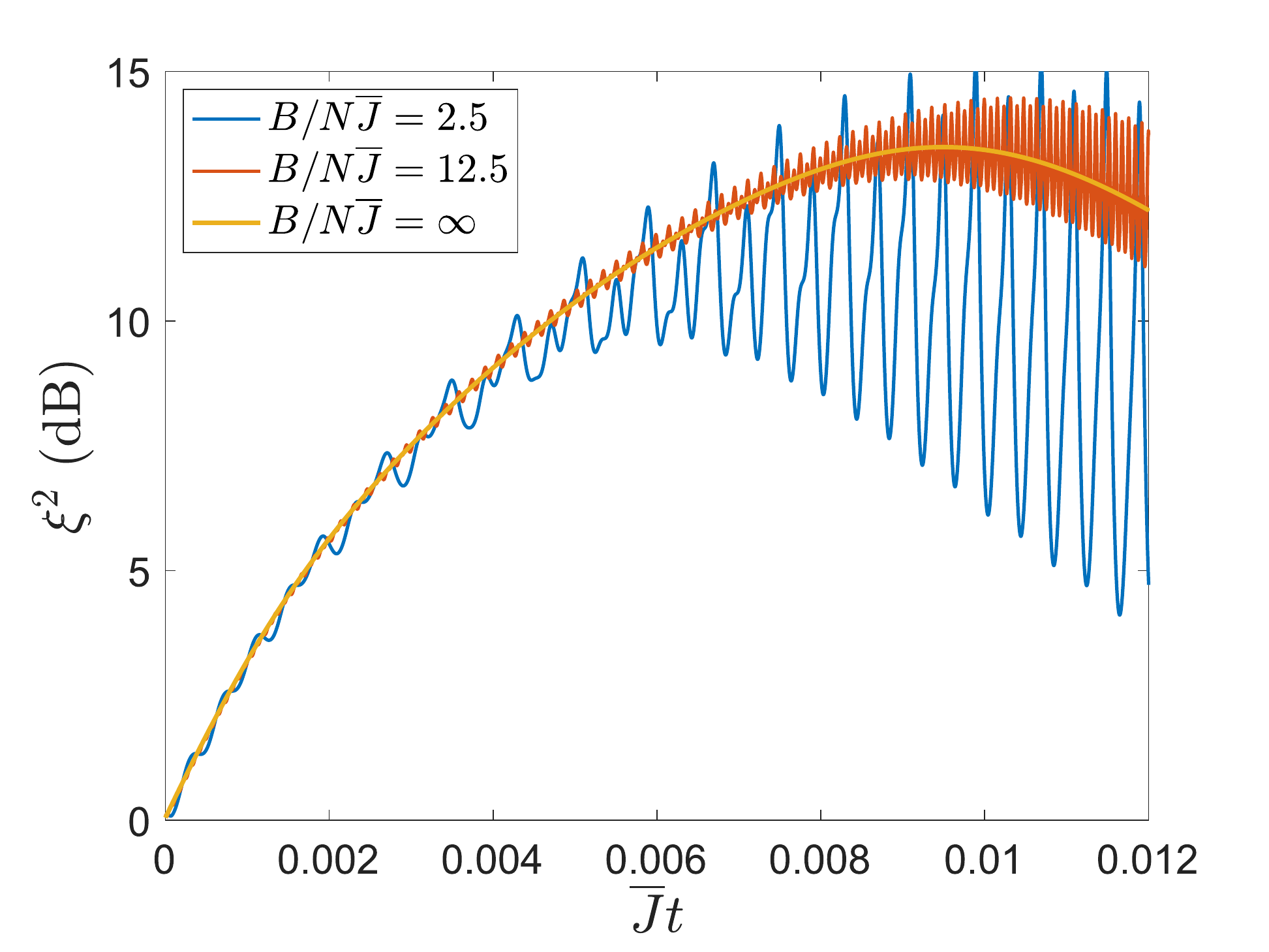}
    \caption{Effect of finite transverse field on the squeezing for (a) $r_b = 2$ and (b) $r_b = 4$. Squeezing is numerically calculated using DTWA with $10^4$ samples in a $14 \times 14$ lattice.}
    \label{fig:finiteB}
\end{figure}

In Fig.~\ref{fig:finiteB}, we investigate the squeezing in a $14 \times 14$ lattice with open boundary conditions and a vdW tail for blockade radii of $r_b=2,4$ and for $B/\overline{J} = 2.5, 12.5$, and the limit of infinite transverse field where the RWA is valid. We see that when there is a finite transverse field, there are oscillations in the squeezing. As the transverse field is increased, these oscillations increase in frequency and decrease in magnitude. Note that each oscillation corresponds approximately to half a Rabi cycle, so the squeezing is maximal when the Bloch vector is near either of the two poles of the Bloch sphere. Interestingly, we see that at these maxima, the squeezing for the finite transverse field can potentially exceed that of the infinite transverse field. While this requires stopping the evolution at the right time, we note that for $B/N \overline{J} = 2.5$, the optimal squeezing time corresponds to approximately 12 Rabi cycles, so for transverse fields of the order of $10-100$ kHz, stopping the evolution near one of the maxima is feasible. Note that as the transverse field is increased, the number of Rabi cycles will increase accordingly.

\section{Benchmarking DTWA in 1D}
\label{bench}
\begin{figure}[h!]
    \centering
    \includegraphics[scale=.6]{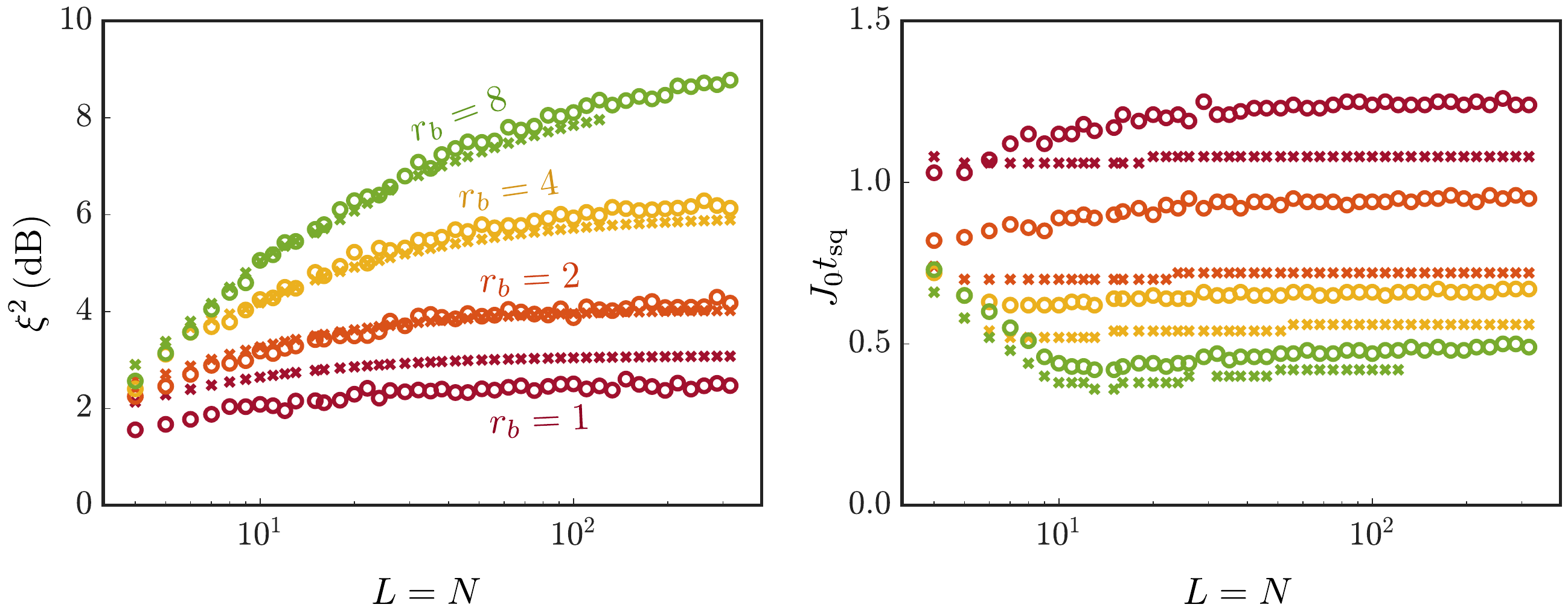}
    \caption{Benchmarking DTWA in 1D. Comparison between TDVP (x's) and DTWA (o's) results for the optimal squeezing, in decibels (left), and the corresponding squeezing time, scaled by the nearest-neighbor coupling $J_0$ (right), obtained with the XX model as given by Eq.~(2) in the main text for various system lengths $L$ and potential ranges $r_b$. We utilize a potential with a sharp cutoff, as well as open boundary conditions for various chain lengths $L = N$. DTWA results are averaged over 10,000 stochastic trajectories. For our TDVP results, our we utilize a time step $J_0 dt = 0.02$, resulting in the discrete jumps observed in the corresponding values of $J_0 t_{\textrm{sq}}$.}
    \label{fig:TDVP}
\end{figure}

In this section, we benchmark the discrete truncated Wigner approximation (DTWA) (see \cite{Schachenmayer2015a,Schachenmayer2015b,Zhu2019}, and also Sec.~\ref{DissDTWA} below) by comparing to results based on time-evolved matrix product states (MPS) \cite{Schollwock2011}. Owing to the generic difficulty of simulating the exact dynamics of higher-dimensional systems or systems exhibiting interactions between distant spins with MPS, we benchmark DTWA for 1D chains, utilizing a sharp cutoff in the potential with no power-law tail and open boundary conditions. We utilize  the time-dependent variational principle (TDVP) to provide quasi-exact solutions to the time-dependent Schr{\"o}dinger equation for our MPS \cite{Haegeman2011,Haegeman2016,Wall2012,Jaschke2018}. In Fig.~\ref{fig:TDVP}, we compare results for the optimal squeezing generated by the XX model (see Eq.~(2) in the main text) in 1D. We generally find improved agreement for both the predicted squeezing and the time at which this occurs as the potential range increases, and the system becomes increasingly connected. In fact, for all $r_b > 1$ shown, we observe excellent agreement in the predicted amount of attainable squeezing. While we average these DTWA results over 10,000 trajectories, in the main text we utilize 20,000 trajectories for Fig.~2(a,b), and 40,000 trajectories for Fig.~3.

For smaller $r_b$ where the most notable discrepancies arise, we observe that DTWA underestimates the attainable squeezing, as well as overestimates the time at which this occurs. Thus, in our examination of the effects of decoherence in Fig.~3, where the dynamics at longer times are increasingly susceptible to the degrading effects of the finite Rydberg lifetime, we expect that DTWA provides, at worst, a conservative estimate for the attainable squeezing. Overall however, our benchmarking suggests that such deviations, when they occur, should remain small. Furthermore, both the consideration of higher-dimensions (i.e. 2D) and the addition of a power-law tail lead to enhanced connectivity of our lattice, and we expect this to lead to further improvement of our results. In fact, similar benchmarks in power-law interacting systems in 2D demonstrate that DTWA yields reliable results for the spin squeezing dynamics \cite{Perlin2020,Muleady2022}.

\section{Experimental parameters}
\label{exp}
In this section, we shall discuss the experimental parameters used to produce Fig.~3 of the main text. First, we note that the we restrict the range of the parameters such that $\Omega \leq 2 \pi \times \SI{10}{\MHz}$ and $N \overline{J} \leq 2 \pi \times \SI{20}{\kHz}$, where the latter restriction is to ensure the validity of the RWA. We considered three commonly-studied Rydberg $S$ (i.e., zero angular momentum) atoms: $^{133}$Cs, $^{87}$Rb, and triplet $^{88}$Sr. To extract the interaction strengths and decay rates, we utilize the ARC code \cite{Robertson2021}. To determine the lattice spacing, there are two behaviors we take into account. First, we ensure that there are no significant level crossings due to the interactions that would lead to a weak Rydberg interaction, weakening the blockade effect. Second, we ensure that the dipole-dipole interactions are perturbative at twice the lattice spacing. Here, we take this to be the point at which the two-atom eigenstate is 95\% $|ss\rangle$, i.e., 5\% of the eigenstate involves other Rydberg states. Although this implies that the eigenstate at shorter distances is strongly composed of additional Rydberg states, for most of the blockade radii considered in Fig.~3, this is well into the blockaded region, so as long as the effective Rydberg blockade interaction is not significantly reduced, the soft-core potential will not be strongly modified. 

In the case of $^{88}$Sr, the numerical methods for extracting the decay rates are inaccurate since it is an alkaline-earth atom. In this case, we rely on experimentally-measured values. In particular, we use Ref.~\cite{Kunze1993} to extract the spontaneous emission rate's scaling behavior $\gamma_{\text{se}} = a n^{*-3}$, where $n^*$ is the effective principle quantum number, for $n=19-23$. To incorporate the effect of blackbody radiation, we utilize measurements from an ongoing experiment at $n = 41$ that is consistent with a lifetime of at least \SI{20}{\micro\second} \cite{KaufmanExpt} and fit the total decay rate $\gamma = a n^{*-3} + b n^{*-2}$ to extrapolate to arbitrary $n^*$ at $T = 300$ K, where we have utilized the fact that the blackbody radiation rate will scale approximately as $n^{*2}$. Fitting the experimental values, we find $a = \SI{2070}{\micro\second^{-1}}$, $b = \SI{15.8}{\micro\second^{-1}}$. For $n=80$ in the main text, this corresponds to a lifetime of \SI{137}{\micro\second}.

\begin{table}[b]
\renewcommand{\arraystretch}{1.3}
    \centering
    \begin{tabular}{c|c|c|c}
         & $a$ & $C_6/2 \pi$ & $\tau$ \\ \hline
         $^{88}$Sr $41^3$S$_1$ & \SI{0.651}{\micro\metre} & \SI{1.5}{\GHz\ \micro\metre^6} & \SI{20}{\micro\second} \\ \hline 
         $^{88}$Sr $60^3$S$_1$ & \SI{1.79}{\micro\metre} & \SI{156}{\GHz\ \micro\metre^6} &  \SI{61.3}{\micro\second}  \\ \hline 
         $^{88}$Sr $80^3$S$_1$ & \SI{3.76}{\micro\metre} & \SI{4.8}{\THz\ \micro\metre^6} &  \SI{137}{\micro\second}  \\ \hline 
         $^{87}$Rb 60S & \SI{1.74}{\micro\metre}& \SI{138}{\GHz\ \micro\metre^6} & \SI{101}{\micro\second}  \\ \hline 
         $^{133}$Cs 60S & \SI{1.62}{\micro\metre} & \SI{107}{\GHz\ \micro\metre^6} & \SI{95.6}{\micro\second}  \\ 
    \end{tabular}
    \caption{Lattice spacings $a$, vdW dispersion coefficients $C_6$, and lifetimes $\tau$ at 300 K used in Fig.~3 of the main text.}
    \label{tab:exptab}
\end{table}

The lattice spacing $a$, $C_6$, and lifetimes for the states considered in Fig.~3 of the main text are listed in Table \ref{tab:exptab}. Although we have focused on a particular set of Rydberg states, we can determine how the behavior changes for different $n$ through scaling arguments. First, we note that the energy difference between different Rydberg states scales like $n^{*-3}$, while the dipole-dipole interaction dispersion coefficient scales like $C_3 \propto n^{*4}$. The first of these two scaling behaviors implies that we should take $\Omega, \Delta \propto n^{*-3}$, implying $J_0 \propto n^{*-3}$. Additionally, in order for the dipole-dipole interactions to continue to remain perturbative, the dipole-dipole interactions must scale with the Rydberg state energy differences, i.e., $C_3/a^3 \propto n^{*-3}$, which implies $a \propto n^{*7/3}$. Since the vdW dispersion coefficient scales like $C_6 \propto n^{*11}$, we see that the vdW interactions scale like $C_6/a^6 \propto n^{*-3}$. As a result, the blockade radius (in units of the lattice spacing) does not scale with $n^*$, and $N\overline{J} \propto n^{*-3}$, which implies smaller transverse fields are needed with increasing $n^*$. Moreover, we see that $\gamma/N \overline{J} \propto a+b n^*$, so the presence of blackbody radiation leads to worse decoherence with increasing $n^*$, although based on the values of $a,b$ for $^{88}$Sr above, we see that the effect is relatively small even at room temperature.

\begin{figure}
    \centering
    \includegraphics[scale=.55]{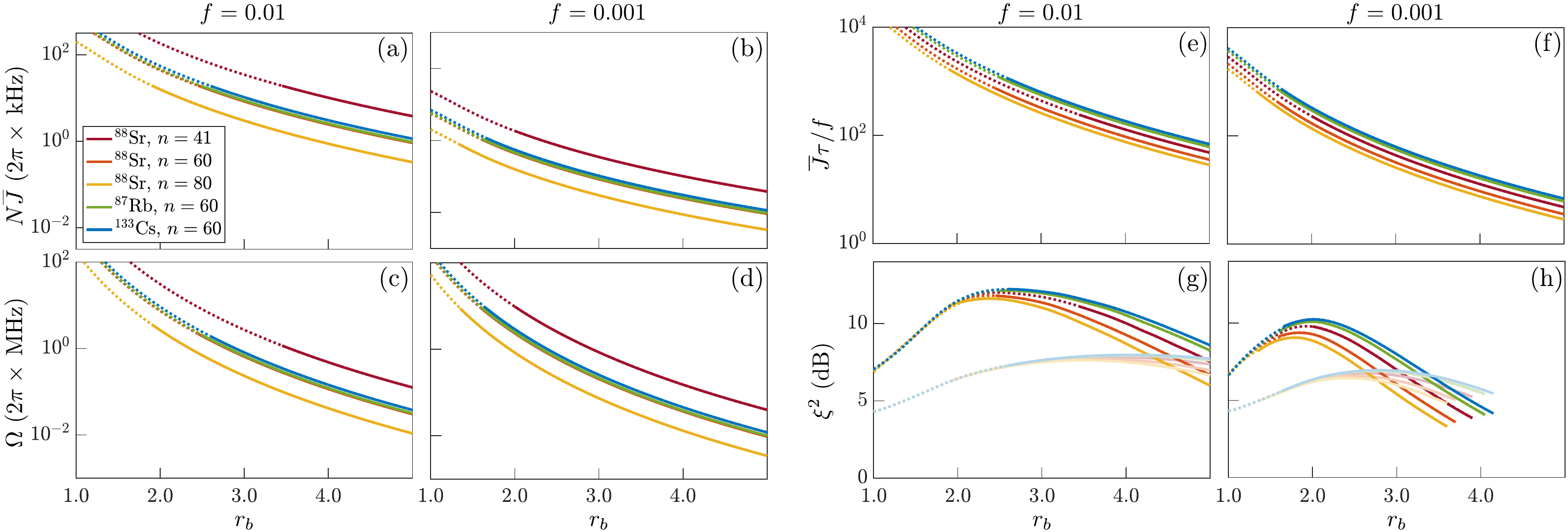}
    \caption{Parameters used for the overlaid curves in Fig.~3 of the main text as a function of $r_b$, in addition to the achievable spin squeezing. We plot the values of (a-b) $N\overline{J}$, (c-d) $\Omega$, and (e-f) $\overline{J}\tau/ f$ for each atom considered. The dotted continuations of each line denote parameters for the maximum cutoff on $\Omega$ ($2\pi\times 10$ MHz) or $N\overline{J}$ ($2\pi\times 20$ kHz) has been exceeded. (g-h) We also show the spin squeezing achievable for these parameters, and compare to the corresponding Ising dynamics, denoted by the faded lines. For the Ising results, we only consider a cutoff on $\Omega$, as we do note require a restriction on $N\overline{J}$ for implementing a transverse field. We show results for Rydberg fractions $f=0.01$ (a,c,e,g) and $f=0.001$ (b,d,f,h). Note that the values of $N\overline{J}$ and $\Omega$ for $^{88}$Sr and $^{87}$Rb with $n=60$ lie virtually on top of each other.}
    \label{fig:Fig3_params}
\end{figure}

In Fig.~\ref{fig:Fig3_params}, we plot the values of $N\overline{J}$ and $\Omega$ as well as the dimensionless quantity $\overline{J}\tau/f$ as a function of $r_b$ for each curve shown in Fig.~3, corresponding to each Rydberg level and atom considered, and different fixed Rydberg fractions $f$; we additionally consider results for $^{88}$Sr with $n=60$, not shown in the main text. We also plot the corresponding value of the spin squeezing in the presence of the relevant decoherence for the considered atom, Rydberg level, Rydberg fraction $f$, and blockade radius $r_b$. Values of $r_b$ for which either $\Omega$ or $N\overline{J}$ exceeds the restricted range, i.e. $\Omega > 2\pi\times 10$ MHz or $N\overline{J} > 2\pi\times 20$ kHz are denoted by a dotted line; in Fig.~3, we simply terminate the curves when these conditions are violated. For the comparable Ising results shown in Fig.~\ref{fig:Fig3_params} (which we do not plot in Fig.~3), we only impose the restriction on the value of $\Omega$, since we do not need to implement a transverse field for this protocol. We note that for $f=0.01$, where the associated interaction timescales are typically large compared to when $f=0.001$, the threshold on $N\overline{J}$ is exceeded before the threshold on $\Omega$ for the chosen examples, whereas the opposite tends to be the case with $f=0.001$. As also evident in Fig.~3 of the main text, we observe that for fixed $f$, an increase in $r_b$ is also accompanied by a decrease in $N\overline{J}$, and the dynamics become increasingly susceptible to the effects of decoherence, leading to a comparably faster degradation of the achievable spin squeezing for the XX model vs the Ising model.

\section{Dissipative DTWA}
\label{DissDTWA}
To treat incoherent processes in the system, we use the semiclassical dissipative discrete truncated Wigner approximation (DDTWA) \cite{Huber2022,Barberena2022} to simulate the dynamics of the master equation in Eq.~(5a) of the main text. We formulate a semiclassical description of our system, introducing classical variables $\mathcal{S}_i^\mu$ corresponding to the value of $s_i^\mu$, where $\mu = x,y,z$ and $1 \leq i \leq N$. For an initial spin-polarized state along $+z$, we form a discrete Wigner function
\begin{align}
    W(\vec{\mathcal{S}}_i) = \frac{1}{4}\Big[\delta(\mathcal{S}_i^x - 1/2) + \delta(\mathcal{S}_i^x + 1/2)\Big]\Big[\delta(\mathcal{S}_i^y - 1/2) + \delta(\mathcal{S}_i^y + 1/2)\Big]\delta(\mathcal{S}_i^z - 1/2).
\end{align}
For each spin, this amounts to the four phase space points $(\mathcal{S}_i^x,\mathcal{S}_i^y,\mathcal{S}_i^z) = (\pm 0.5, \pm 0.5, 0.5)$ each occurring with equal probability $1/4$. The coherent dynamics are then obtained by solving the associated classical equations of motion of the relevant Hamiltonian, in conjunction with randomly sampling initial values for $(\mathcal{S}_i^x,\mathcal{S}_i^y,\mathcal{S}_i^z)_{1\leq i\leq N}$ according to the above distribution. Incoherent terms in our master equation may be accounted for by the addition of stochastic noise terms to our classical equations of motion. For further details, see \cite{Gardiner2009,Huber2022,Barberena2022}.

For an ensemble of dynamical trajectories with initial conditions sampled from our initial Wigner distributions, quantum expectation values may then be approximated via $\langle s_i^\mu(t)\rangle \approx \overline{\mathcal{S}_i^\mu(t)}$, where $\overline{\,\cdot\,}$ denotes averaging with respect to this ensemble. Likewise, symmetrically-ordered correlators may be obtained via $\langle(s_i^\mu s_j^\nu + s_j^\nu s_i^\mu)(t)\rangle/2 \approx \overline{\mathcal{S}_i^\mu(t)\mathcal{S}_j^\nu(t)}$. Given the generic nonlinear nature of our classical equations of motion, this averaging produces results beyond mean-field theory that take into account the effect of the quantum noise distribution on the dynamics \cite{Schachenmayer2015a,Schachenmayer2015b,Zhu2019}.   

\bibliography{RydbergSqueezing,Extra}